\documentclass[aps,prl,twocolumn,showpacs,%
preprintnumbers,amsmath,superscriptaddress,floatfix]{revtex4}

\usepackage{graphicx}
\usepackage{dcolumn}
\usepackage{bm}

\begin{document}
\title{Topological nature of polarization and charge pumping in ferroelectrics}
\author{Shigeki Onoda}
\affiliation{
Tokura Spin SuperStructure Project, ERATO,
Japan Science and Technology Agency,\\
c/o Department of Applied Physics, University of Tokyo, Tokyo 113-8656, Japan
}
\author{Shuichi Murakami}
\affiliation{
CREST, Department of Applied Physics, University of Tokyo, Tokyo 113-8656, Japan
}
\author{Naoto Nagaosa}
\affiliation{
Tokura Spin SuperStructure Project, ERATO,
Japan Science and Technology Agency,\\
c/o Department of Applied Physics, University of Tokyo, Tokyo 113-8656, Japan
}
\affiliation{
CREST, Department of Applied Physics, University of Tokyo, Tokyo 113-8656, Japan
}
\affiliation{
Correlated Electron Research Center, 
AIST, Tsukuba, Ibaraki 305-8562, Japan
}
\date{\today}
\begin{abstract}
Electric polarization in insulators is represented by 
the transferred charge through a shift of the Bloch wavefunctions
induced by an adiabatic change of external parameters $\vec{Q}$. 
It is found that this covalent/quantum contribution 
is determined nonlocally by the topological structure in the 
${\vec Q}$ space. The condition for the charge pumping 
for a cyclic change of $\vec{Q}$ is also obtained.
Applications of this picture to various organic ferroelectrics and BaTiO$_3$
are discussed.
\end{abstract}
\maketitle

Dielectric properties of insulators have been among the most important 
subjects in condensed-matter physics \cite{ash}. 
For insulators, each electron is often regarded as ``localized'', 
and is treated as a point charge. 
Hence the classical picture of polarization, i.e., the displacement times 
the charge, demands an ionic nature for ferroelectricity.
On the other hand, it has recently been recognized that 
covalency of the electronic state also contributes 
to ferroelectricity \cite{ash,resta}. 
For this mechanism, extended Bloch wavefunctions in the crystal are essential.
An interference pattern of Bloch wavefunctions 
is sensitive to a change of parameters such as atomic displacements, 
leading to a shift of the electrons as a whole \cite{resta}. 

Experimentally, several novel quasi-one-dimensional organic ferroelectrics have been found \cite{ni,tmttf,horiuchi}.
In these ferroelectrics, the polarization is mostly perpendicular 
to displacements of atoms or molecules, and is much larger 
than that expected from the classical picture. In particular, 
some of hydrogen-bonded organic co-crystals with neutral molecules show 
a large polarization arising from a slight change of the hydrogen position, 
which can never be understood in the classical picture \cite{horiuchi}.
This motivates us to study one-dimensional models.

In one dimension, the covalent contribution to the polarization 
$dP$ induced by an adiabatic change of 
external control parameters $\vec{Q}$ is expressed as \cite{resta,thouless}
\begin{eqnarray}
&& dP = |e|\vec{A}\cdot d\vec{Q},
\label{eq:dP}\\
&& A_\alpha=-2\sum_{\nu}
\int\frac{dk}{2\pi}
n_\nu(k)
\text{Im}
\left\langle
\frac{\partial \phi_{\nu}}{\partial k}\right|
\left.
\frac{\partial \phi_{\nu}}{\partial Q_\alpha}\right\rangle,
\label{eq:A}
\end{eqnarray}
where $k$ is a wave number, and $|\phi_{\nu}(k)\rangle$ and 
$n_\nu(k)$ represent an eigenstate and a momentum distribution 
function for the band (including the spins) $\nu$, respectively. 
$A_\alpha$ is a linear response of the covalent part of polarization, 
and a generalization of the Born charge tensor.
The choice of $\vec{Q}$ is at our disposal, such as atomic displacements, electric field, and pressure.
Equation~(\ref{eq:A}) involves the Berry phase \cite{berry,shapere} of the Bloch wavefunctions in the momentum space. Here one can see similarity to the quantum Hall effect \cite{tknn} and anomalous Hall effect \cite{onoda}.

In contrast to the ionic contribution, 
it is difficult to have an intuitive picture for the electronic contribution 
arising from the covalency. Only the sophisticated first-principle 
band calculations have succeeded in estimating the polarization \cite{resta}.
Thus, it is highly desirable to explore a quantum and intuitive 
picture of polarization in covalent ferroelectrics 
and to provide a guiding principle for a design of such ferroelectrics, which is a main purpose of this Letter.
The basic idea is to consider the space of external control parameters 
${\vec Q}$, instead of the momentum space which has been focused in the previous works~\cite{tknn,onoda,monopole}. 
This reveals a hidden topological structure in the $\vec{Q}$ space, 
expressed in the form of the magnetostatics in this space.
We show that a trajectory of band-crossing $\vec{Q}$ points  
constitutes a ``circuit'' with a quantized total ``current''.
The vector field ${\vec A}({\vec Q})$ in Eq.~(\ref{eq:A})
then corresponds to the ``magnetic field'' generated by the circuit. 
We can calculate a change $\Delta P$ in the polarization due to that of 
$\vec{Q}$ as a contour integral of $\vec{A}(\vec{Q})$.

This framework is related to the quantized charge pumping
discussed by Thouless \cite{thouless} in the following way.
When the parameters $\vec{Q}$ are changed along a cycle $C$ enclosing 
the ``circuit'', an integer multiple of the charge $e$ is transferred 
from one end to the other of the sample, corresponding to the charge 
pumping. On the other hand, the usual polarization  
corresponds to a fraction of charge pumping where the 
$C$ does not form a loop. This picture also facilitates to  
design insulators showing the quantum charge pumping.

We take a one-dimensional two-band Hamiltonian;
\begin{eqnarray}
  {\cal H} &=& \sum_{k,\sigma,\sigma'}
  c_{k,\sigma }^\dagger
  H_{\sigma\sigma'}(k,\vec{Q})
  c_{k,\sigma'},
  \label{eq:calH}\\
  H(k,\vec{Q})&=&\epsilon_{0}(k,\vec{Q})+
  \sum_{\alpha=1,2,3}h_\alpha(k,\vec{Q})\sigma_\alpha,
  \label{eq:H}
\end{eqnarray}
where $c_{k,\sigma}$ and $c_{k,\sigma}^\dagger$ represent the annihilation and creation operators of an electron of a wave number $k$ and an internal degree of freedom $\sigma$, which will be identified as orbitals later.
$\sigma_{\alpha}$ ($\alpha=1,2,3$) are the Pauli matrices.  
We have omitted the spin degrees of freedom, 
which can be taken into account by multiplying $\Delta P$ by two. Though it is a simplified model, 
it contains essential features of the nontrivial topology.
Even in one spatial dimension, by regarding the parameters $Q_{\alpha}$ 
as additional dimensions, the topological structure becomes equivalent to that of higher dimensional systems, and can be nontrivial.
The eigenenergies are given by $\varepsilon_\pm(k,\vec{Q}) = \epsilon_{0}(k,\vec{Q}) \pm h(k,\vec{Q})$ with $h\equiv|\vec{h}|$. Below we study geometrical and topological properties of $\vec{A}$ in {\it insulating states} where the chemical potential $\mu$ lies in the gap between the two bands; $|\epsilon_0(k,\vec{Q})-\mu|\le h(k,\vec{Q})$ for any $k$.

\begin{figure}[htb]
\begin{center}\leavevmode
\includegraphics[width=6cm]{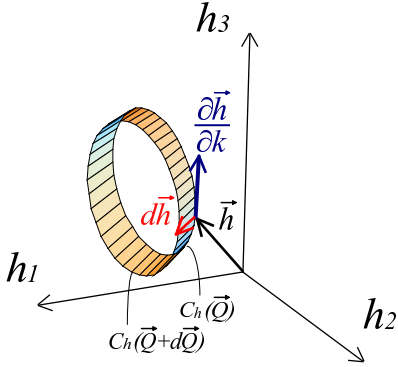}
\end{center}
\caption{Change of polarization induced by an adiabatic change of $\vec{Q}$ is schematically shown in the $\vec{Q}$ space.}
\label{fig:A-Contour}
\end{figure}

Equation~(\ref{eq:A}) is then written in terms of $\vec{h}$ as
\begin{eqnarray}
A_\alpha &=& 
-
\int_{-\pi/a}^{\pi/a}\frac{dk}{4\pi}
\hat{h}\cdot\left(\frac{d\hat{h}}{dk}\times
\frac{d\hat{h}}{dQ_\alpha}\right),
\label{eq:Aa}
\end{eqnarray}
where $\hat{h}=\vec{h}/h$, and $a$ is the lattice constant. Equation (\ref{eq:Aa}) gives a geometrical meaning to the polarization; an infinitestimal change $d P$ is given by a solid angle of a ribbon in the $\vec{h}$ space bounded by the two loops which are drawn by $\vec{h}(k,\vec{Q})$ and $\vec{h}(k, \vec{Q} + d\vec{Q})$ for $-\pi/a \leq k \leq \pi/a$ (Fig.~\ref{fig:A-Contour}). Therefore $d P$ is maximized when a shift of the loop is ``perpendicular'' to the loop itself.

\begin{figure}[htb]
\begin{center}\leavevmode
\includegraphics[width=5cm]{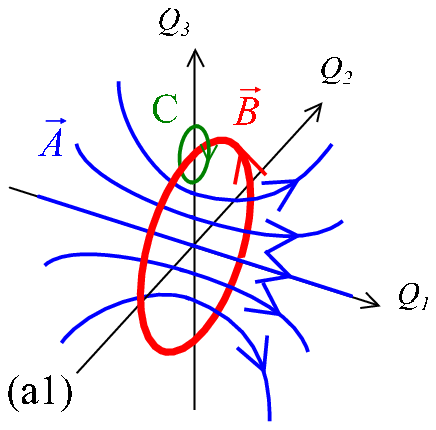}
\includegraphics[width=3.3cm]{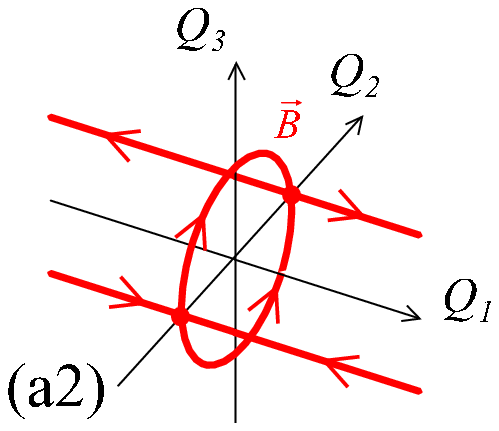}
\includegraphics[width=3.3cm]{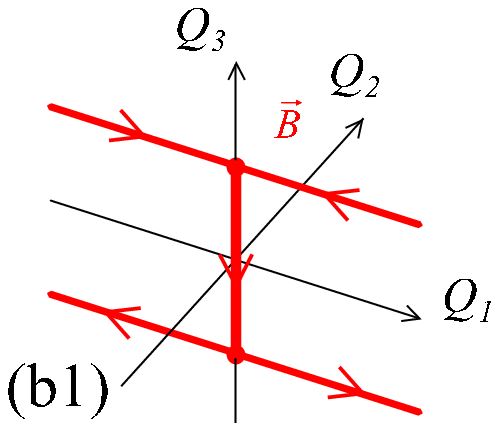}
\includegraphics[width=3.3cm]{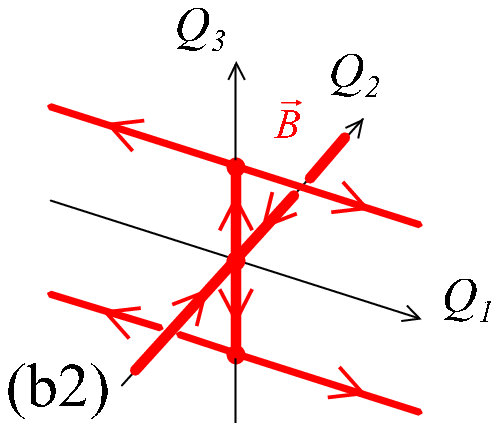}
\end{center}
\caption{Geometrically different string structures of $\vec{B}$ (red) in the $\vec{Q}$ space for different classes of models in the text. In (a1), we also show the vector field $\vec{A}$. The loop $C$ encompasses the flux string. The arrows on the flux strings show the directions of $\vec{B}$ in the strings. Total flux of each string is unity, except for the strings on the $Q_3$ axis in (b1) and on the $Q_2$ and the $Q_3$ axes in (b2), whose total flux is two.}
\label{fig:string}
\end{figure}

Henceforth, we assume that $\vec{Q}$ has three components $\vec{Q}=(Q_1, Q_2, Q_3)$, with which we can use an analogy to magnetostatics; 
generalization to cases with many components is immediate.
To investigate topological properties of $\vec{A}$ in the $\vec{Q}$ space, we introduce a ``flux density'' $\vec{B}=\vec{\nabla}_Q\times\vec{A}$ from Eq.~(\ref{eq:Aa}), namely,
$B_\alpha=
-
\frac{3}{4}\int_{-\pi/a}^{\pi/a}\frac{dk}{2\pi}
\epsilon_{\alpha\beta\gamma}
\frac{d\hat{h}}{dQ_{\beta}}\cdot\left(\frac{d\hat{h}}{dk}\times
\frac{d\hat{h}}{dQ_{\gamma}}\right)$.
In general, $\vec{B}(\vec{Q})$ has the following three properties. (i) $\vec{B}$ is nonzero only along strings corresponding to band crossings $h(k,\vec{Q})=0$. Actually, it shows a $\delta$-function singularity there. (ii) $\vec{\nabla}_{Q}\cdot\vec{B}(\vec{Q})=0$, and hence the string forms a loop or goes from infinity to infinity, as exemplified in Fig.~\ref{fig:string}. (iii) The total flux of the string is quantized to be an integer. In the rest of this paragraph, we derive these properties.
Since $|\hat{h}|=1$, three vectors $\frac{d\hat{h}}{dQ_{\beta}}$, $\frac{d\hat{h}}{dk}$, and $\frac{d\hat{h}}{dQ_{\gamma}}$ are all perpendicular to $\hat{h}$ and hence are coplanar. This leads to $\vec{B}(\vec{Q})=\vec{0}$. However, this does not apply for band-crossing $\vec{Q}$ points where there exists $k$ satisfying $h(k,\vec{Q})=0$. Such band-crossing $\vec{Q}$ points form a curve, i.e., a ``string'' in the $\vec{Q}$ space. This leads to (i). (iii) is shown by considering the total flux of the strings surrounded by a loop $C$,
\begin{equation}
\int_C d\vec{Q}\cdot\vec{A}(\vec{Q})
= \int_S d\vec{S}_{Q}\cdot\vec{B}(\vec{Q})
  \label{eq:P-surface}
\end{equation}
where $S$ is the closed surface surrounded by the loop $C$. 
It corresponds to the wrapping number of the unit sphere with the periodic boundary condition both for $k$ and ${\vec Q}$ directions. Namely, Eq.~(\ref{eq:P-surface}) is the Pontryagin index for the mapping from $(k,\vec{Q})$ ($\vec{Q}$ is on the loop $C$) to $\vec{h}$, and is always an integer. (ii) follows because $\vec{A}(\vec{Q})$ is a continuous physical observable with no gauge degrees of freedom and contains a singularity only on the strings. Practical calculations of $\vec{B}$ are explained in \cite{B}.

The property (iii) is related with the quantized charge pumping discussed by Thouless \cite{thouless};
when $\vec{Q}$ is changed along a closed cycle $C$ within the insulating states, the amount of pumped charge is the integer (\ref{eq:P-surface}). 
It represents the linking number between $C$ and the strings.

Next we consider an asymptotic behavior of $\vec{A}(\vec{Q})$ near the string. Asymptotically, it is inversely proportional to the distance from the string, i.e., to the band gap. From Eq.~(\ref{eq:A}), it can be shown that the direction of $\vec{A}(\vec{Q})$ near the string is the same as that of a ``magnetic field'' due to an ``electric current''. Details will be presented in \cite{longpaper}.

It is useful to decompose $\vec{A}(\vec{Q})$ into transverse and longitudinal parts, $\vec{A}(\vec{Q}) = \vec{A}^t(\vec{Q})+\vec{A}^l(\vec{Q})$, where $\vec{\nabla}_{Q}\times\vec{A}^{l}=0$ and $\vec{\nabla}_{Q}\cdot\vec{A}^{t}=0$. The relation of $\vec{B}(\vec{Q})$ to the transverse component $\vec{A}^t(\vec{Q})$ is analogous to that of current to magnetic field in the electromagnetism. Hence, $\vec{A}^t(\vec{Q})$ can be constructed from $\vec{B}(\vec{Q})$ by the Biot-Savart law. In particular, when the model has a form $\vec{h}(k,\vec{Q})=\vec{f}(k)+\vec{Q}$, the longitudinal component $\vec{A}^l(\vec{Q})$ vanishes and $\vec{A}(\vec{Q})$ is completely described by the Biot-Savart law;
namely, the dielectric response $\vec{A}$ of the system
is determined by the string structure representing the band crossing. 

\begin{figure}[htb]
\begin{center}\leavevmode
\includegraphics[width=6cm]{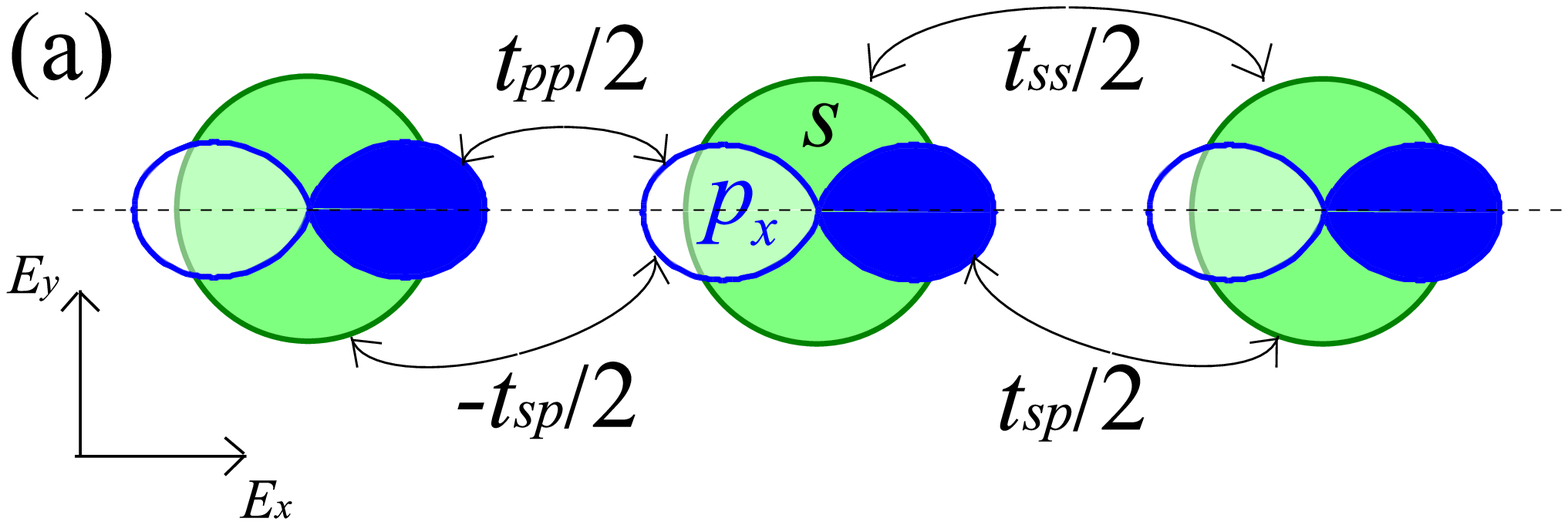}
\includegraphics[width=6cm]{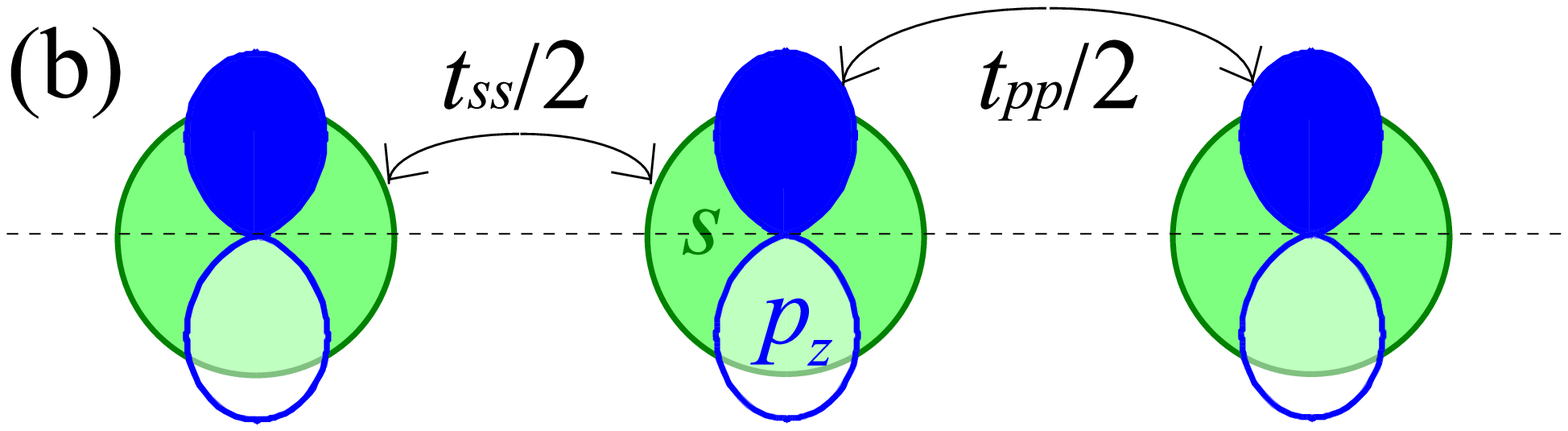}
\end{center}
\caption{One-dimensional two-band models for molecular orbitals are 
schematically shown.}
\label{fig:model}
\end{figure}

Let us show how this theory is applied in specific examples. 
One simplest model contains $s$-like and $p_x$-like orbitals on a molecule, 
aligned along the $x$ direction (see Fig.~\ref{fig:model} (a)). 
This is the simplest model for the hydrogen-bonded organic ferroelectrics 
\cite{horiuchi}. 
Let us introduce three transfer integrals $t_{ss}$, $t_{pp}$, and 
$t_{sp}$, and the Hamiltonian is given by
$H(k,{\vec Q}) = [(t_{ss} + t_{pp})/2] \cos k +
(t_{sp}\sin k) \sigma_2 + ( [(t_{ss} - t_{pp})/2] \cos k )
\sigma_3 + {\vec Q} \cdot {\vec \sigma} $ (see Fig.~\ref{fig:model} (a)).
The corresponding string and the distribution of $\vec{A}(\vec{Q})$ is given 
in Fig.~\ref{fig:string} (a1).
For example, $Q_1$ is the atomic displacement which breaks 
the inversion symmetry, such as the hydrogen atom position 
in hydrogen-bonded ferroelectrics \cite{horiuchi}, 
or an external electric field $E_x$ along the chain direction.
The parameter $Q_{3}$ represents an energy difference between 
$s$ and $p_{x}$ orbitals: $Q_3^{(0)} = (\epsilon_s - \epsilon_p)/2$.
We can modify $Q_3$ by external means. 
For example, the pressure $p$ can tune this energy difference,
i.e., $Q_3 = Q_3^{(0)} + a_p p$ ($a_p$: constant),
which can be of the order of $10\text{ meV}$ with $p \sim 10 \text{ kbar}$ 
\cite{horiuchi}.
The square of the external electric field $E_z$ perpendicular to the 
chain also acts as $Q_3$, because the $s$ and $p_x$ orbitals have 
different polarizability. 
The string intercepts the $Q_{3}$ axis 
at $Q_{3}=\pm \Delta$, where $2\Delta=|t_{ss}-t_{pp}|$ is a gap 
at $k=0$ without the orbital energy difference.


As is evident from Fig.~\ref{fig:string} (a1), $Q_1$, being perpendicular 
to the string, is most effective to produce the polarization. 
The cases $Q_{3}^{(0)}>\Delta$ and $Q_{3}^{(0)}<\Delta$ correspond to
ionic and covalent states, respectively. From the Biot-Savart law, 
in the ionic state the polarization $P$ decreases rapidly, as 
$P \sim ea (t^2 Q_1/(Q_{3}^{(0)})^{3})$
($t \sim t_{ss},t_{pp},t_{sp}$, $a$: lattice constant) 
for $Q_3^{(0)} \gg \Delta$
since the contributions from the segments in the circuit cancels partially, 
whereas in the covalent state ($Q_{3}^{(0)}<\Delta$) 
the polarization $P$ is constant for small $Q_{3}^{(0)}$: $P\sim ea (Q_1/t)$.

Let us discuss possibility of the charge pumping. As described above, 
the electric field and/or the pressure act as 
$Q_1 = a_1 E_x$, $Q_3 = Q_3^{(0)}+ a_3 E_z^2$ or
$Q_3 = Q_3^{(0)}+ a_p p$ ($a_1$, $a_3$, $a_p$: constants).
Therefore, a cycle $C$ in Fig.~\ref{fig:string} (a1), namely a cyclic change 
of $E_x$ and $E_z$ (or $p$) leads to the charge pumping. 
Hence, to achieve this quantum charge pumping, the gap at $k=0$, 
$|Q_3^{(0)}\pm \Delta|$, should be small enough to be comparable with 
the experimentally attainable values of $a_3  E_{z}^2$ or $a_p p$.
Therefore it is promising if we can design an insulating molecular crystal 
with a gap at $k=0$ is $\sim 10 \text{ meV}$.

One can also consider a change of the transfer integrals due to distortion. 
This leads to the term $(Q_1 \cos k) \sigma_1$ instead of $Q_1 \sigma_1$.
With this change, the topology of the string also changes 
into Fig.~\ref{fig:string} (a2).

Next, we consider the $s$- and $p_z$-orbitals (Fig.~\ref{fig:model} (b)). 
The model Hamiltonian is
$H(k,{\vec Q}) = [(t_{ss} + t_{pp})/2] \cos k +
( [(t_{ss} - t_{pp})/2] \cos k + \Delta )
\sigma_3 + Q_1 \sigma_1 +  (Q_2 \sin k)\sigma_2 + Q_3 \sigma_3 $.
where $Q_1 \propto E_z$ mixes the $s$ and $p_z$ orbitals, $Q_2$ represents 
the rotation angle of the $p_z$ orbital, inducing the transfer between the 
$s$ and $p_z$ orbitals, and $Q_3 \propto E_x^2$ changes the difference 
between these two orbitals. The string for this model is shown in 
Fig.~\ref{fig:string} (b1). Instead of $Q_1 \sigma_1$ term in the above 
Hamiltonian, one can introduce a transfer integral between $s$ and $p_z$ 
orbitals by breaking the symmetry between $z$ and $-z$ parts of the 
electron cloud. This leads to $(Q_1\cos k)\sigma_1$, and the resulting 
string is shown in Fig.~\ref{fig:string} (b2).
In both cases, the polarization response will be larger 
if the change of $\vec{Q}$ is along the direction around the strings.

\begin{figure}[htb]
\begin{center}\leavevmode
\includegraphics[width=5.cm]{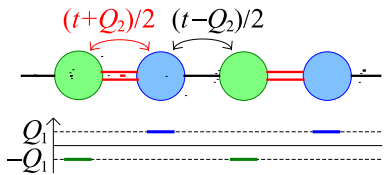}
\end{center}
\caption{An ionic or charge-ordering dimer model for TTF-CA or TMTTF }
\label{fig:dimer}
\end{figure}

Even without two orbitals per one molecules, one-dimensional 
dimerized charge-ordered systems such as TTF-CA and (TMTTF)$_2$PF$_6$ 
gives the same string structure as in Fig.~\ref{fig:string} (b1).
The Hamiltonian reads
\begin{eqnarray}
&&{\cal H} = \sum_i \left[ (t/2)( c_i^\dagger c_{i+1} + h.c.)\right.
\nonumber \\
&& \ \ \ \ \left.
+ Q_1 (-1)^i c_i^\dagger c_i + (Q_2/2) (-1)^i
( c_i^\dagger c_{i+1} + h.c.)\right],
\label{eq:TTFmodel}
\end{eqnarray}
where $Q_1$ represents the charge density order while $Q_2$ the 
dimerization order (see Fig.~\ref{fig:dimer}). 
This Hamiltonian corresponds to 
$H(k,{\vec Q}) =  (t \cos k+ Q_3) \sigma_3 + 
Q_1 \sigma_1 +( Q_2 \sin k) \sigma_2$ with a fictitious $Q_3$,
and the corresponding string is shown in Fig.~\ref{fig:string} (b1). 
For TTF-CA, Eq. (\ref{eq:TTFmodel}) should be regarded as that for one of the 
spin and $Q_1$ is given by the energy difference of the HOMO of donor 
and LUMO of acceptor molecules. In TTF-CA, the molecular displacement 
changes the transfer $Q_2$, giving rise to the polarization almost 
perpendicular to the displacement \cite{ni}. 
In Ref.~\cite{ni}, the electron-phonon interaction, namely $Q_2$, 
is estimated to be $\sim 0.2 \text{eV}$ while the site energy difference is 
$Q_1 \sim 0.4 \text{ eV}$. Because the cyclic change in the $Q_1$-$Q_2$ plane 
around the $Q_3$ axis of Fig.~\ref{fig:string} (b1) transfers the charge $4e$,
one can roughly estimate the polarization $P$ as
$P \cong (2e/\pi a^2)\text{Arctan}(Q_2/Q_1)\cong 0.3 e/a^2$. 
This is consistent with the more elaborated calculation
including the electron correlation \cite{soos}.
(TMTTF)$_2$PF$_6$, on the other hand, is a quarter-filled system, 
and Eq. (\ref{eq:TTFmodel}) should be regarded as the spinless fermion 
model derived for the infinite $U$ limit, describing the charge degrees of 
freedom. In this case, both $Q_2$ and $Q_1$ originate from 
the spontaneous symmetry breaking, which occurs at a low temperature 
\cite{tmttf}. Therefore a sufficiently strong electric field along the 
chain could induce the rotation of $(Q_1,Q_2)$ around the origin,
which is similar to the charge density wave sliding. However, we note 
that the CDW has double periodicity and there is no phase degrees
of freedom.

Ferroelectricity of perovskite oxides like BaTiO$_3$ in the tetragonal phase~\cite{zhong94} can be essentially described in terms of a one-dimensional $d$-$p$ model~\cite{ishihara}, which is identical to Eq.~(\ref{eq:TTFmodel}). The ionic contribution from atomic displacements, which is not included in our tight-binding Hamiltonian, is estimated as follows: the ionic configuration of Ba$^{2+}$, Ti$^{2.89+}$ and O$^{1.63-}$~\cite{cohen} together with each displacement $0.06$ {\AA}, $0.12$ {\AA} and $-0.036$ {\AA} along the chain~\cite{harada} gives a polarization of $P_{\rm ion}\cong0.13e/a^2$. On the other hand, the ionicity leads to a rough estimate of $Q_1/\sqrt{Q_1^2+t^2} \sim 2.89/4$, while according to Ref.~\cite{Harrison}, the $2.4\%$ change of Ti$-$O bond lengths yields an $8.7\%$ alternation of $d$-$p$ transfers, i.e., $Q_2\sim0.087 t$. Therefore the covalent contribution is estimated to be $P_{\rm cov}\cong(4e/\pi a^2) {\rm Arctan}(Q_2/Q_1)\cong0.11e/a^2$, which is comparable to $P_{\rm ion}$. Then, the total polarization is roughly estimated as $P=P_{\rm ion}+P_{\rm cov}\cong0.24e/a^2$, which explains the experimental value $(0.20$-$0.25)e/a^2$~\cite{merz}.

The present argument can be generalized to higher dimensions. 
In three dimensions, for example, by changing $k_y$ and $k_z$,  
the string will form a three-dimensional manifold. In the simplest 
case the string will form a bundle, and $\vec{A}(\vec{Q})$ outside 
this bundle represents the polarization. 
More generic cases are to be discussed elsewhere \cite{longpaper}.

In conclusion, we explored the topological and nonlocal nature of 
polarization in the space of the external parameters ${\vec Q}$. 
The $\vec{Q}$ points where a band crossing occurs play a crucial role 
in determining the polarization of the insulating state.
This clarifies also the condition for the quantized charge pumping,
and will be useful for designing materials with large dielectric response.

We thank R. Shindou, S. Horiuchi, Y. Tokura, and G. Ortiz 
for fruitful discussions. 
This work is supported by Grant-in-Aids from the Ministry of Education,
Culture, Sports, Science and Technology of Japan.


\begin{thebibliography}{99}
\bibitem{B}
{}From Eq.~(\ref{eq:Aa}), the flux of each string, Eq.~(\ref{eq:P-surface}), is an integral on a surface $C\times[-\frac{\pi}{a},\frac{\pi}{a}]$. Via the Gauss' theorem, it reduces to a volume integral over a
three-dimensional region $S\times[-\frac{\pi}{a},\frac{\pi}{a}]$. 
This volume integral is interpreted as a total strength of monopoles 
in this region, where the monopoles are defined by 
$h(k,\vec{Q})=0$. 
The strength of each monopole is $-1$ or $1$, when the Jacobian of the mapping from $(k,\vec{Q})$ ($\vec{Q}$ is on the $S$) to $\vec{h}$ at the monopole is positive or negative, respectively.
\end{thebibliography}
\end{document}